%% file: main.tex
\documentclass[]{eptcs}
\providecommand{\event}{FMAS 2020} % Name of the event you are submitting to
\usepackage{breakurl}             % Not needed if you use pdflatex only.
\usepackage{underscore}           % Only needed if you use pdflatex.
\usepackage{graphicx}
\usepackage[export]{adjustbox} % Allows framing of figures.
\usepackage{url}
\usepackage{listings}
\usepackage{cite}
\usepackage{pgf}
\usepackage{tikz} 
\usepackage{xcolor}

\input{macros}

\title{From Requirements to Autonomous Flight:\\An Overview of the
  Monitoring ICAROUS Project }

\author{ %\hspace*{-20pt}
  \begin{tabular}[t]{ccc} 
Aaron Dutle$^1$  & Laura Titolo$^2$ & Dimitra Giannakopoulou$^3$ \\
    C{\'e}sar Mu{\~n}oz$^1$ & Ivan Perez$^2$ & Anastasia Mavridou$^{4}$ \\
    Esther Conrad$^1$ & Swee Balachandran$^2$ & Thomas Pressburger$^3$\\
    Alwyn Goodloe$^1$ & \quad & \quad \end{tabular}  \\
  {\footnotesize 
    \vspace{-2pt}$^1$NASA Langley Research Center, Hampton, Virginia} \\
   \vspace{-2pt} {\footnotesize  $^2$National Institute of Aerospace, Hampton, Virginia }\\
   \vspace{-2pt}{\footnotesize                                                                  $^3$NASA
                                                                      Ames
                                                                      Research
                                                                      Center,
                                                                      Moffett
                                                                      Field,
                                                                      California }
                                                                      \\ 
     {\footnotesize              $^4$KBR Inc. / NASA Ames Research Center, Moffett
                    Field, California}
                  \\
         \vspace{-2pt} {\textit  {\footnotesize \{aaron.m.dutle, cesar.a.munoz, esther.d.conrad, a.goodloe,  laura.titolo, ivan.perezdominguez,}}\\
     {\textit  {\footnotesize sweewarman.balachandran, dimitra.giannakopoulou, anastasia.mavridou, tom.pressburger\}@nasa.gov}}
 }
  
\def\titlerunning{Monitoring ICAROUS}
\def\authorrunning{A.~Dutle et al.}
\begin{document}
\maketitle

\begin{abstract}
\input{abstract}
% The \emph{Monitoring ICAROUS} project integrates 3 NASA-developed
% products toward the goal of simply specifying safety requirements,
% creating run-time monitors from them, and integrating them into an autonomous
% flight system. \fret is used to specify safety requirements for an aircraft in a structured
% natural language. These requirements are then translated into Copilot
% specifications,
% which are used to generate executable runtime
% monitors. These monitors are then integrated into ICAROUS, an
% autonomous system for carrying out UAS operations.  
\end{abstract}

\section{Introduction}
\label{sect:intro}
\input{introduction}

\section{Tool Descriptions}
\label{sect:tools}

%\subsection{FRET}
%\label{subsect:fret}
\input{fret}
\input{copilot}

%\subsection{ICAROUS}
%\label{subsect:icarous}
\input{icarous}

\section{Integration}
\label{sect:integration}
\input{integration}

\subsection{ICAROUS interface to FRET}
\label{icarous-fret}
\input{icarous-fret}

\subsection{FRET to Copilot translation }
\label{fret-copilot}
\input{fret-copilot}

\subsection{Copilot Monitors in ICAROUS}
\label{copilot-icarous}
\input{copilot-icarous}

\section{Conclusion}
\label{sect:conclusion}
\input{conclusion}

%\nocite{*}
\bibliographystyle{eptcs}
\bibliography{generic}
\end{document}

%% file: abstract.tex
The Independent Configurable Architecture for Reliable Operations of Unmanned Systems\linebreak (ICAROUS)
is a software architecture incorporating a
set of algorithms to enable autonomous operations of unmanned aircraft
applications. This paper provides an overview of {\em Monitoring ICAROUS}, a
project whose objective is to provide a formal approach to generating
runtime monitors for autonomous systems from requirements written in a structured
natural language. This approach integrates \fret{}, a formal
requirement elicitation and authoring tool, and Copilot, a runtime
verification framework. \fret{} is used to specify formal requirements in
structured natural language. These requirements are translated
into temporal logic formulae. Copilot is then used to generate
executable runtime monitors from these temporal logic
specifications. The generated monitors are directly integrated into
ICAROUS to perform runtime verification during flight.

%% file: introduction.tex
% !TEX root = MI.tex

%The safety assurance of autonomous systems is becoming increasingly important in the modern world.
% The automotive industry is already equipping vehicles with up to level 2 (conditional automation) systems such as Tesla's \emph{Autopilot} and Cadillac's \emph{Super Cruise}, and some entrants such as NAVYA and Waymo are demonstrating level 4 (high automation) systems \cite{synopsis20}.
% In the aviation industry, automation of some aviation functions (facilitated by Flight Management Systems and others) has reduced the workload of current pilots, while full-scale autonomous flight is still out of reach.
%One factor contributing to the lack of autonomous aircraft systems is that such systems are highly complex cyber-physical systems (combinations of hardware and software interacting with the physical world), and must have high safety assurance in order to be fielded in the National Airspace.
%A challenge in the development of autonomous aircraft systems is the safety case that enables these  
%highly complex cyber-physical systems to be integrated into the airspace system. 
% In this context,
%NASA's
The Independent Configurable Architecture for Reliable Operations of Unmanned Systems \linebreak
\mbox{(ICAROUS)~\cite{CMHNB2016DASC}} is a software architecture for enabling safe autonomous operation of unmanned aircraft systems (UAS) in the airspace. The primary goal of ICAROUS is to provide autonomy to enable beyond visual line of sight (BVLOS) missions for UAS without the need for constant human supervision/intervention.
ICAROUS provides highly-assured functions to avoid stationary obstacles, maintain a safe distance from other users of the airspace, and compute resolution and recovery maneuvers. 

Hardware and software verification via formal methods offers the highest assurance of safety available for such cyber-physical systems. While  there have been considerable advances in creating industrial-scale 
formal methods (e.g., \cite{Souyris2009,kaivola2009,cook2018}), it is not yet practical to apply them to an entire complex system such as ICAROUS.
Formal verification is generally carried out on a model of a system rather
than the software itself, and so the properties verified may not hold if the
model is inaccurate or if other faults make the system behave unpredictably.
Moreover, while there has been much progress made in verification of neural networks in particular (\cite{Katz2019, Julian2019}, increasingly autonomous systems
employing machine learning and similar methods are challenging for formal
verification.

Runtime verification (RV)~\cite{2008:havelund:verifyyourruns} %,monitors,introRV}
is a verification technique that has the potential to
enable the safe operation of safety-critical systems that are too complex to
formally verify or fully test. In RV, the system is monitored during execution,
and property violations can be detected and acted upon during the mission. RV detects when properties are violated at runtime, so it cannot enforce the correct operation of a system, but is an %significant
improvement over testing alone, and can enhance testing by finding real cases of requirement violation. Copilot~\cite{2020:perez:copilot3} is a runtime verification framework developed by NASA researchers and others. %Additionally, RV is a primary component of the Simplex architecture, which allows a complex, possible untrusted, component to operate until a runtime monitor detects that it has violated a safety property or exited a safety envelope, at which point a simpler, verified component takes over the operation.

While RV can be used to monitor and detect property violations, the actual properties to be monitored must be determined and specified externally. Such safety requirements are generally written by hand in natural language, which can lead to ambiguity as to their meaning or applicability.
%In addition, to be used in runtime monitoring, even a clear requirement must be translated into executable code that checks the property.
Additionally, when runtime verification is used as a key safety component of an autonomous system, having clearly specified requirements that are properly translated into \emph{executable} monitors is critical. \fretish{}~\cite{GiannakopoulouP20} is a structured natural language developed by NASA to write unambiguous requirements.
The associated tool, \fret{}~(Formal Requirements Elicitation Tool), provides a framework to write, formalize and analyze requirements and automatically generate temporal logic formulae from them.

The \emph{Monitoring ICAROUS} project, a work-in-progress joint effort at NASA, will demonstrate the integration of robust requirements-based runtime verification applied to an autonomous flight system for unmanned aircraft using \fret{}, Copilot, and ICAROUS.
This project brings together work that the NASA formal methods team has been doing for many years on requirements elicitation and specification, runtime verification, and assured autonomous aircraft software.
% 
%The integrated runtime monitors will be automatically generated from a set of requirements written in the \fretish{} language.
%\fretish{}\cite{GiannakopoulouP20a,GiannakopoulouP20} is a structured natural language developed by NASA Ames to write unambiguous requirements.
% 
%The associated tool, \fret{}~(Formal Requirements Elicitation Tool), provides a framework to write, formalize and analyze requirements.
%These requirements can be specified by the user either through requirement templates, or through an interactive interface.
%\fret{} automatically generates temporal logic formulas that can be used by verification and analysis tools.
% The NASA Ames team have developed \fret{} (the Formal Requirements Elicitation Tool) and the \fretish{} language to address the requirements specification problem. 
% 
%These formal requirements become input to Copilot, a framework for specifying runtime monitors, and are translated into C code.

% Prior to this effort, these three topics have been developed at NASA independently. 
% Finally, ICAROUS is a software system for autonomous aircraft maintained by NASA Langley, built on top of NASA's long-standing core Flight System (cFS) component-based flight software architecture.

\begin{figure}[h]
  \centering
\includegraphics[width = 0.8 \linewidth, frame]{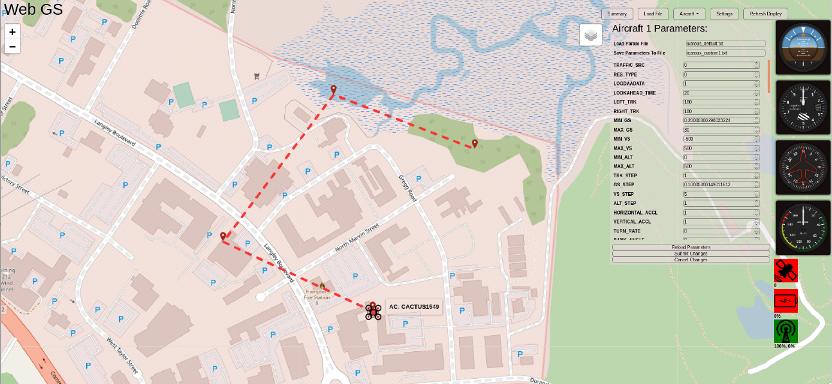}
\caption{The current interface for setting parameters in ICAROUS.}
\label{fig:icarous-settings}
\end{figure}

The concept of operation for the integrated system is simple. Prior to the start of an autonomous flight with ICAROUS, an operator can set a collection of different mission and safety parameters (see Figure~\ref{fig:icarous-settings}). For example, the detect-and-avoid module can be specified to avoid other aircraft by at least 250 ft horizontally and 50 ft vertically. % and the geofencing module can be given polygons around buildings or an area that the aircraft must avoid by at least 50 ft.
With integrated monitoring, the related implicit requirements will become explicit ones, expressed in the structured natural language of \fretish{}, and available to be viewed and edited by the user. In addition, a method for specifying custom requirements similar to the \fret{} interface will be available. These requirements expressed in \fretish{} will be translated into Copilot's monitoring language, which will generate C code for the RV monitors. These monitors will be integrated into ICAROUS, which will use them to determine requirements violations. A violation of a monitor will alert the operator, who can use the information to return the aircraft to a safe state. %, either through an issued command or by taking manual control.

As a motivating example, the remainder of the paper will use the detect and avoid requirement \emph{``Requirement 1: While flying, remain separated from an intruder aircraft by at least 250 ft horizontally or 50 ft vertically.''} This safety property will be followed through the chain of tools employed, and illustrate the work to be done in the integration. %Section~\ref{sect:tools} will describe in more detail each of the three tools in the project, including how the example requirement would be used. Section~\ref{sect:integration} will discuss the work being undertaken to integrate the three, as well as some related tasks on verification of parts of the integrated system. Section~\ref{sect:conclusion} will examine possible extensions of the project, discuss related work, and conclude.    

%% file: fret.tex
\fret{}\footnote{\url{https://github.com/NASA-SW-VnV/fret}} %\cite{GiannakopoulouP20, GiannakopoulouP20a}
is an open-source tool developed at NASA for writing, understanding, formalizing, and analyzing
requirements. %\footnote{The description here only touches a few basic capabilities of \fret{} applicable to the \emph{Monitoring ICAROUS} project, the interested reader is directed to \cite{GiannakopoulouP20, GiannakopoulouP20a} for more thorough treatments.}
In practice, requirements are typically written in natural language, which is ambiguous and, consequently, not
amenable to formal analysis. Since formal, mathematical notations are unintuitive, requirements in \fret{} are
entered in a restricted natural language named \fretish{}. \fret{} helps users write \fretish{} requirements, both
by providing grammar information and examples during editing, but also through textual and diagrammatic
explanations to clarify subtle semantic issues.

\begin{figure}
  \centering
    \includegraphics[width = 0.8 \linewidth, frame]{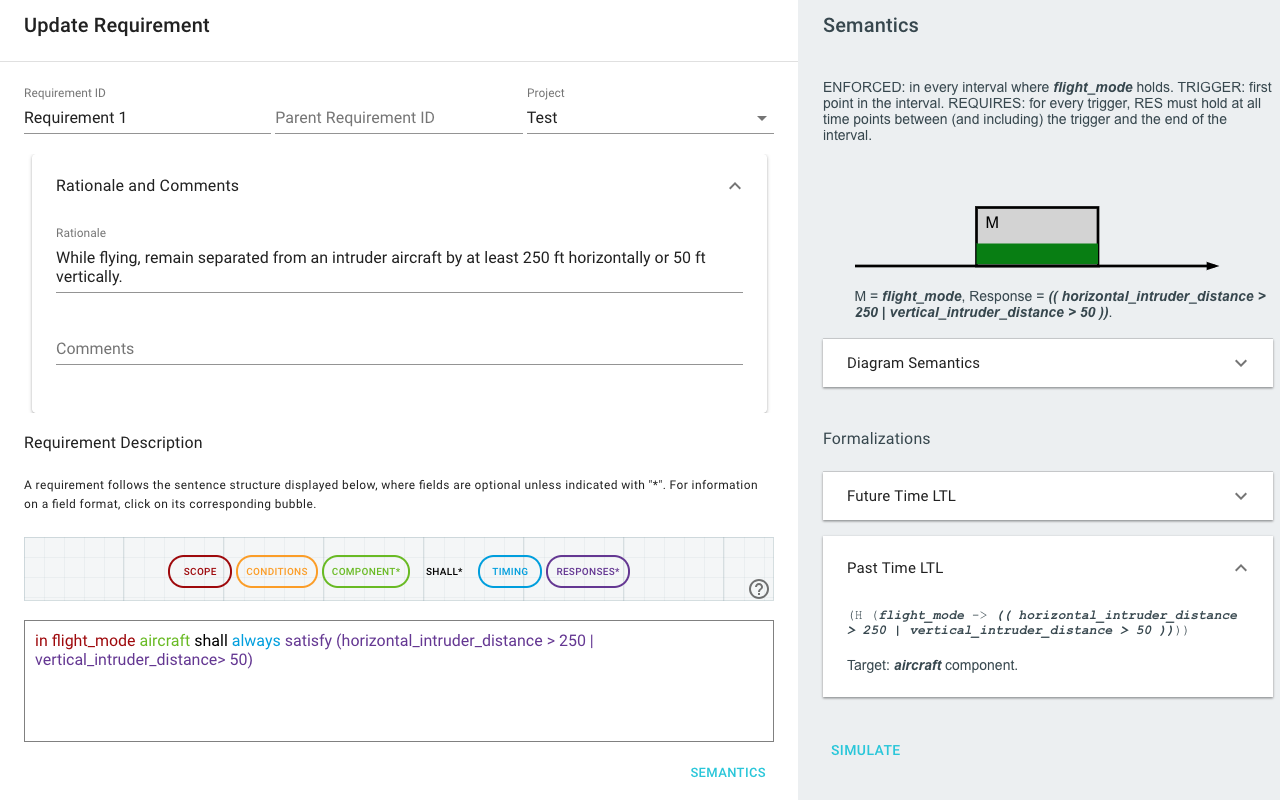}
    
\caption{\fret{} editor, with the example requirement entered, and the \emph{Semantics} pane visible.}
 \label{fig:fret-req}
  \end{figure}

Figure \ref{fig:fret-req} illustrates \fret{}'s requirements elicitation interface, with the example \emph{Requirement 1} entered. The ``Rationale and Comments'' field holds the original text requirement; the ``Requirement Description'' field is where the \fretish{} requirement is composed. Once a requirement is entered, the ``Semantics'' pane shows a text description of the \fretish{} requirement, displays a ``semantic diagram'' showing a visual explanation of the requirement applicability over time, and provides translations from \fretish{} to Metric Future- and Past-time linear temporal logic (LTL) \cite{Koymans90}.

A \fretish{} requirement description is automatically parsed into six sequential fields; \emph{scope,
condition, component, shall, timing,} and \emph{response}, with the \fret{} editor
dynamically coloring the text corresponding to each field (Fig.~\ref{fig:fret-req}). %Help and examples on each specific field can be displayed
%in the help tab by clicking on the corresponding field bubble.
The mandatory \emph{component} field specifies the
component that the requirement applies to (\textbf{aircraft}). The \emph{shall} keyword states that the component
behavior must conform to the requirement. The \emph{response} field currently is of the form \emph{satisfy R}, where \emph{R} is a
non-temporal Boolean-valued expression (\textbf{horizontal_intruder_distance$>$250 $|$ vertical_intruder_distance$>$50}).
The (optional) \emph{scope} field states that the requirement is only assessed during particular modes,
for the example, in \textbf{flight_mode}. The (optional) Boolean expression field \emph{condition} states that,
within the specified mode, the requirement becomes relevant only from the point where the condition becomes
true. The (optional) \emph{timing} field specifies
at which points the response must occur, in the example ``always'', meaning at all points in flight mode.

\fret{} automatically produces formulas in several formal languages, including Metric Future-time LTL and Past-time LTL.
% An extensive verification framework ensures that the generated formulas conform to the semantics of the \fretish{} language. 
%
% In addition to the visualization provided by the semantic diagram,
\fret{} also offers an interactive visualizer, 
%available by clicking \textsc{simulate} in the semantics view of the help tab. Given a \fret{}
%requirement, it
showing temporal traces of each of the signals (variables) involved as well as the valuation of the
requirement for each point in time. %(see Figure 4).

%% file: copilot.tex
Copilot\footnote{\url{https://copilot-language.github.io/}} %~\cite{2020:perez:copilot3,2010:pike:copilot},
is an open-source runtime verification framework for real-time embedded
systems. %, initially developed by Galois, Inc. and the National Institute of
%Aerospace under contract to NASA. 
%
Copilot monitors are written in a compositional, stream-based language. The
framework translates monitor specifications into C99 code with no dynamic
memory allocation and executes with predictable memory and time, crucial in
resource-constrained environments, embedded systems, and safety-critical
systems.

The Copilot language has been designed to be high-level, easy to understand,
and robust. To prevent errors in the RV system that could affect systems
during a mission, the language uses advanced programming features to provide
additional compile-time and runtime guarantees. For example, all arrays in
Copilot have fixed length, which makes it possible for the system to detect,
before the mission, some array accesses that would be out of bounds. 

Copilot supports a number of logical formalisms for writing
specifications including  
a bounded version of Future-time Linear Temporal Logic
\cite{1977:pnueli}, Past-time Linear Temporal Logic (PTLTL) \cite{LaroussinieMS02}, and  Metric
Temporal Logic (MTL) \cite{Koymans90}. Copilot also includes support libraries
with functions such as majority vote, used to implement
fault-tolerant monitors~\cite{2013:pike}.
\emph{Requirement 1} expressed as  a PTLTL Copilot 
specification is as follows: 
\vspace{-0.2cm}
\begin{center}
\begin{lstlisting}[language=Haskell,basicstyle=\small\ttfamily]
alwaysBeen (flightMode ==> (  horizontalIntruderDistance > 250
                           || verticalIntruderDistance   > 50 ))
\end{lstlisting}
\end{center}

% In the context of this project, we are extending FRET to automatically produce
% a Copilot monitor for each requirement, either from the FRETISH variant of the
% past-time LRL expressions, or from their CoCoSpec counterparts. With this
% connection in place, requirements can be further translated into C functions
% that check that the property specified by the requirement holds during runtime.
% We are designing this integration to take place transparently for the user:
% they would write FRETISH requirements using FRET on one end, click a button or
% run a script provided with our tool, and obtain C code on the other end. The
% final C files can be integrated in a larger application, such as one that uses
% Core Flight System to monitor incoming signals from an Autonomous Vehicle.

%% file: icarous.tex
ICAROUS\footnote{\url{https://github.com/nasa/icarous}}
%\cite{CMHNB2016DASC, BMCFPDASC2018}
is an open-source software
architecture developed at NASA to enable safety-centric
autonomous aircraft missions.
%\footnote{This section only briefly
                         %describes ICAROUS. The interested reader is
                         %referred to \cite{CMHNB2016DASC,
                         %BMCFPDASC2018} for a more thorough treatment
                         %of the capabilities and services provided by
                         %ICAROUS.}.
It is a service-oriented architecture, where service applications provide
various capabilities such as path planning,
sense and avoid, geofence containment, task planning, and more, through a publish-subscribe middleware. 
Applications are logically
organized into conflict detectors, conflict resolvers, mission
managers, and decision makers. 

Conflict detectors are algorithms that check for imminent
violation of constraints such as geofences, conflicts
due to other vehicles in the airspace, deviations from mission
flight plan, etc. These conflict detecting applications can
also provide \emph{tactical} resolutions, which provide a simple maneuver which
will prevent the corresponding conflict. Conflict resolvers compute resolutions to prevent imminent
violation of specified constraints. Resolvers may 
handle multiple conflicts simultaneously, and can provide
\emph{strategic} resolutions that are computed to prevent one or
more constraint violations. A decision making application receives conflict information
from monitors and triggers resolvers to compute
resolutions for one or more conflicts. When resolving imminent
constraint violation, outputs from mission applications are ignored.
The mission is resumed once all conflicts are
resolved.

These services are connected through the
NASA core Flight System (cFS) middleware,\footnote{\url{https://cfs.gsfc.nasa.gov/}} a platform-independent
reusable software framework and a set of reusable software
applications. The three key aspects to the cFS architecture are
a dynamic run-time environment, layered software,
and a component-based design. These key aspects make
cFS suitable for reuse on any number of embedded software
systems. The cFS middleware simplifies
the flight software development process by providing the
underlying infrastructure and hosting a runtime environment
for development of project/mission specific applications.

%The addition of runtime monitors to ICAROUS is intended to provide a higher, system-level, non-interfering monitoring capability that the conflict monitors don't provide. Conflict monitors are meant to work on individual conflicts, detect them before occurrence, and provide tactical solutions. This means that conflict monitors are generally complicated algorithms, while also being restricted to a particular subsystem. Runtime monitors are meant to detemine if global requirements are violated, thus can assess system level properties, while being simpler due to the fact that they only assess violations, not predict or tactically resolve them.   

%% file: integration.tex
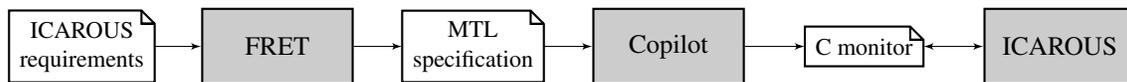
\begin{figure}[h]
\begin{center}
\begin{tikzpicture}[auto, node distance=2cm,>=latex']
    \node [doc] (fretishSpec) {ICAROUS\\ requirements};
    % \node [doc, above of = fpProg, yshift=-0.5cm] (realProg) {PVS Real\\Specification};
    % \node [doc, below of=realProg, yshift=0.5cm] (ranges) {Input Ranges};    
    \node [block, right of= fretishSpec,  xshift=0.6cm, minimum height=1cm, minimum width=2cm, font= \small, fill=mygray] (fret) {FRET};
    \node [doc, right of= fret, xshift=0.6cm] (mtlSpec) {MTL\\specification};
    \node [block, right of= mtlSpec,  xshift=0.6cm, minimum height=1cm, minimum width=2cm, font= \small, fill=mygray] (copilot) {Copilot};
    \node [doc, right of= copilot, xshift=0.6cm] (monitor) {C monitor};
    \node [block, right of= monitor, xshift=0.6cm, minimum height=1cm, minimum width=2cm, font= \small, fill=mygray] (icarous) {ICAROUS};
    % % \node [block, below of= precisa, yshift=0.2cm, xshift=0.8cm,  font= \footnotesize] (fprock)  {FPRoCK};
    % \node [doc, right of = precisa, xshift=1.2cm] (pvscert) {PVS round-off errors\\ certificates};
    % \node [block, below of= pvscert, yshift=0.5cm, minimum height=1cm, minimum width=2cm, font= \small] (pvs) {PVS};
    % \node [doc, right of = precisa, xshift=1.2cm, yshift=1.5cm]  (ccode) {instrumented\\ACSL/C program};
    % \node [block, right of= ccode,  xshift=1.5cm, minimum height=1cm, minimum width=2cm, font= \small] (framac) {Frama-C};
    % \node [doc, below of= framac, minimum height=1cm, minimum width=2cm] (VCs){Verification\\Conditions}; 

    \draw [->] (fretishSpec) -- node[] {} (fret);
    \draw [->] (fret) -- node[] {} (mtlSpec);
    \draw [->] (mtlSpec) -- node[] {} (copilot);
    \draw [->] (copilot) -- node[] {} (monitor);
    \draw [<->] (monitor) -- node[] {} (icarous);
    % % \draw [->] (precisa) -- node[] {$\tauProgDecl{}$} (fpProg);
    % \draw [->] (precisa) -- node[] {} (pvscert);
    % \draw [->] (pvscert) -- node[] {} (pvs);
    % \draw [<->, dashed] (precisa) -- (kodiak) ;
    % \draw [->] (realProg) -- node[] {} (precisa);
    % \draw [->] (ccode) -- node[] {} (framac);
    % \draw [->] (precisa) -- node[] {} (ccode);
    % \draw [->] (framac) -- node[] {} (VCs);
    % \draw [->] (VCs) -- node[] {} (pvs);
\end{tikzpicture}
\end{center}
\caption{Toolchain to automatically generate monitors for ICAROUS.}
    \label{fig:diagram}
\end{figure}

The integration of the three systems described in Section \ref{sect:tools} requires several steps in order to create a complete framework. The three major steps in the integration are as follows. The first step %(see Subsection \ref{subsect:icarous-fret})
is developing a method for ICAROUS-specific requirements and safety properties to be specified in \fret{}. Next%(Subsection \ref{subsect:fret-copilot})
, the requirements expressed in \fretish{} must be translated into Copilot. Finally%(Subsection \ref{subsect:copilot-icarous})
, the monitors generated by Copilot must be integrated into ICAROUS in
a usable way. Each of these integration steps are discussed in turn
below, and the toolchain is depicted in Figure~\ref{fig:diagram}.

%% file: icarous-fret.tex
In order to facilitate the use of \fret{} for ICAROUS-specific requirements, several things need to be done. The first of these is a thorough accounting of all of the modes, systems, and signals available within ICAROUS for a monitor to access. 

In \fret{}, the specification of safety properties is completely independent of the system that is being considered. This makes \fret{} very general and powerful in its ability to craft requirements, but makes it somewhat cumbersome to use in the specific setting of ICAROUS. The variables that \fret{} uses to refer to modes, systems, and signals in requirements are completely arbitrary, while being able to use such a requirement as a monitor means that each of these variables must correspond to an actual system or signal in ICAROUS. After each of these possible variables in ICAROUS is identified, along with its datatype information, \fret{} can be restricted to using only information that can be obtained by the system in specifying properties. 

Another task to be completed is the generation of ICAROUS-specific
templates for \fret{}. Many of the safety requirements that an autonomous
flight system will be expected to follow exhibit similar patterns from
flight to flight. For example, an autonomous operation may be required
to stay below a certain altitude ceiling. More specific requirements
such as \emph{Requirement 1} are parametric requirements that a detect-and-avoid (DAA) system is expected to obey based on settings in the
DAA system. For such requirements, a template can be given allowing
the user to input particular values based on the mission, and the
associated requirement added to those to be monitored.  Additionally,
many of the existing settings in an ICAROUS configuration carry with
them implicit safety requirements. Setting the \emph{max_altitude}
parameter in the DAA system of ICAROUS can be interpreted as a
requirement that the aircraft never goes above the value set. Parsing
these files can allow for the \emph{automatic} generation of a
requirement from the associated template and the variable setting.

%Finally, the most visible of the tasks to be done for this part of the integration is the creation of an interface allowing for a novice user to be able to create, alter, and generate monitors from specified requirements. The goal is to have something similar to Figure \ref{fig:icarous-settings}. In this interface, a drop-down menu would allow the creation of a requirement from a list of templates or written by hand, a button would allow the generation of default requirements from the existing system settings, a different view would allow for the current requirements to be viewed and edited, and a button would finalize and generate monitors from the requirements. One technical issue with this vision is that the current interface (fig.~\ref{fig:icarous-settings}) alters settings in the actual deployed instance of ICAROUS onboard the aircraft. In order to include code for monitors in ICAROUS, these requirements must be finalized  and monitor code incorporated \emph{before} ICAROUS is loaded onboard.   

%% file: fret-copilot.tex
The main integration step between \fret{} and Copilot is to take a requirement
expressed in \fretish{}, and translate it into a Copilot monitor. A prototype
tool named Ogma %\footnote{Ogma is often considered a god of speech and
%language in Irish and Scottish mythology}
is being developed to create Copilot
monitors from languages such as \fretish{}, SPEAR~\cite{2017:fifarek:spear2}, and
AGREE~\cite{2012:cofer:agree}. The tool can translate the Past-time 
LTL formulas \fret{} generates, so it can process any requirements specified in
\fretish{}.  The resulting Copilot monitor can then be automatically translated
into C99 code that checks that a corresponding property holds during runtime.
This conversion should occur transparently: users specify \fretish{}
requirements, and automatically obtain C code compatible with ICAROUS.

%% file: copilot-icarous.tex
The integration of Copilot-generated monitors into ICAROUS should be fairly straight-forward. Since ICAROUS is already a service-oriented publish/subscribe architecture, the main issue is subscribing and routing the appropriate signals to the monitors, and returning a signal to the user that indicates which properties have been violated. To  facilitate this integration, the Ogma tool automatically generates a cFS application, responsible for subscribing to the appropriate ICAROUS applications, making data available to Copilot, and handle runtime violations reported by Copilot.

A technical difficulty in this approach (mentioned in Section~\ref{icarous-fret}) is that the RV monitors are generated \emph{after} requirements are specified, and then this C code is integrated into ICAROUS. Currently, ICAROUS is installed on the aircraft once, and, generally, only settings are changed between flights. With new monitors generated for each flight, ICAROUS must be configured, new monitors generated, and then the system installed on the aircraft before flight. A partial solution would be to include parametric versions of common monitors, and have the parameters instantiated prior to flight. Alternatively, a utility for including monitoring code could be added, since the RV service would not change.

%% file: conclusion.tex
% !TEX root = MI.tex
The work described here is still in-progress. Additional work is being conducted tangential to this integration. The \fretish{} language and semantics are being specified in the Prototype Verification System (PVS)~\cite{pvs}, to prove that the evaluation of a \fretish{} statement is the same as the evaluation of the LTL statement produced by \fret{} for all possible finite and infinite traces. Currently, the verification of the translation is done through a systematic, rigorous testing framework for finite traces up to a certain length. An embedding of \fretish{} in PVS would also allow for formal reasoning about models of a system such as ICAROUS with respect to specified requirements.

Related work on
% for each of the three separate aspects of
 requirements specification, RV, and autonomous flight systems is omitted here. The interested reader is directed to the corresponding sections of \cite{GiannakopoulouP20, GiannakopoulouP20a} for requirements, \cite{2020:perez:copilot3, 2010:pike:copilot, 2008:havelund:verifyyourruns,introRV} for runtime verification, and \cite{CMHNB2016DASC, BMCFPDASC2018} for autonomous flight systems.
Work that has a similar flavor to the integration of these tools includes \cite{CauwelsHHJR20}, where the R2U2~\cite{SchumannMR15} engine is used to monitor an automated and intelligent UAS Traffic Management System for adherence to safety requirements during operation.
The specifications are written in the Mission-time Linear Temporal Logic (MLTL)~\cite{ReinbacherRS14}, an extension of MTL, in contrast with the present approach where the specifications are given using structured natural language. %VeriMon+\cite{BasinDHKR0T20} is a formally verified monitor for Metric First-Order Temporal Logic (MFOTL~\cite{BasinKMZ15}). 

%The \emph{Monitoring ICAROUS} project will allow a user to define relevant requirements automatically or using the structured natural language \fretish{}, have these requirements automatically translated into runtime monitors using Copilot, and have these monitors integrated seamlessly into the autonomous flight system ICAROUS. The framework supports simple requirements specification and analysis, and robust runtime verification, with the translation and integration steps performed in the background. Requirements-based runtime monitoring demonstrates a real-world application of formal methods to increase the safety assurance of complex automated systems.

The \emph{Monitoring ICAROUS} project will allow a user to obtain relevant requirements automatically or defined using the structured natural language \fretish{}, translate these requirements into runtime monitors using Copilot, and seamlessly integrate these monitors into the autonomous flight system ICAROUS. The framework supports simple requirements specification and analysis, with robust runtime verification, while the translation and integration steps are performed in the background. Requirements-based runtime monitoring demonstrates a real-world application of formal methods to increase the safety assurance of complex automated systems.